\documentstyle[12pt]{article}
\input{psfig}

\begin{document}
\textheight 23cm
\textwidth 16cm
\oddsidemargin 0in
\evensidemargin 0in
\topmargin -0.25in

\centerline{\bf \Large Integer and fractional charge solitons in}
\centerline{\bf \Large modulated strips in the fractional quantum}
\centerline{\bf \Large Hall regime}
\vskip .5cm
\centerline{\large E. Ben-Jacob$^1$, F. Guinea$^2$, 
Z. Hermon$^3$ and A. Shnirman$^1$}
\vskip.5cm
$^1$ School of Physics and Astronomy, Tel Aviv University,
69978 Tel Aviv, Israel. \\
$^2$ Instituto de Ciencia de Materiales,
CSIC, Cantoblanco, E-28049 Madrid, Spain. \\
$^3$Institut f\"ur Theoretische Festk\"orperphysik,
Universit\"at Karlsruhe, D-76128 Karlsruhe, Germany.

\begin{abstract}
We propose the existence and study the solitonic excitations in two
kinds of samples in the fractional quantum Hall regime. One is a
strip modulated by a one-dimensional array of gates. The other is
made of two parallel strips coupled by a one-dimensional
array of tunnel barriers. We predict the existence of integer charge
solitons in the first case, and fractional charge solitons in the
second case. We study the two cases both in the dissipative and in the
inertial limits.
\end{abstract}
\vskip 0.2cm

\section{Introduction}

\begin{figure}
\centerline{\psfig{figure=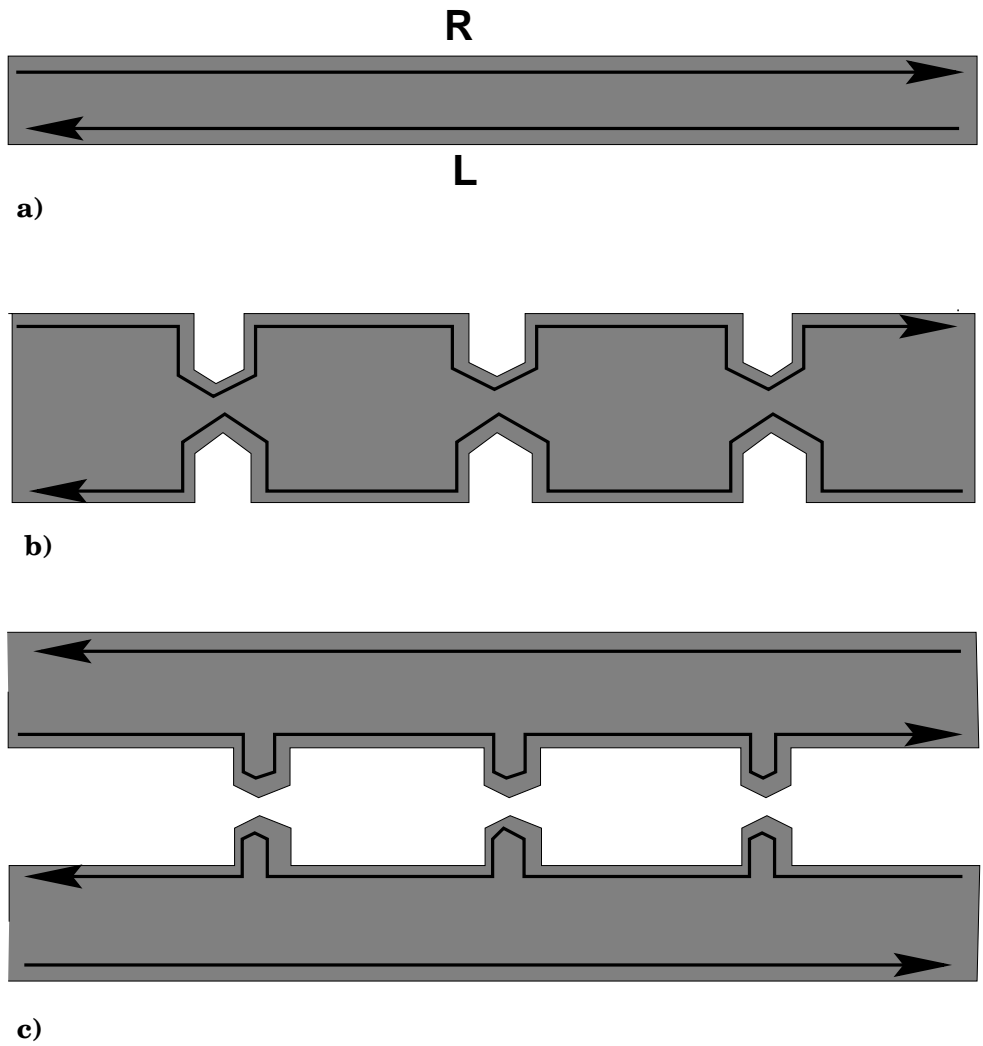,width=13.0cm}}
\caption{a) The right and left edge currents;
b) System I: FQH strip with gates;
c) System II: Two FQH strips connected by tunnel barriers.}
\label{FIG_1}
\end{figure}

Samples in the fractional quantum Hall (FQH) regime have been studied 
extensively, since they are in a novel quantum state and exhibit an 
interesting spectrum of excitations \cite{Prange_Girvin}. 
Here we predict the existence and 
study the properties of solitonic excitations in such
systems. Motivated by the 
investigations of fluxons in quasi one-dimensional (1D) 
long Josephson junctions 
\cite{Josephson} and charge solitons in 1D arrays of normal \cite{normal} and 
Josephson \cite{arrays} junctions, we consider the two systems shown in 
Fig. (\ref{FIG_1}). 

Both systems are made of strips (quasi 1D) of samples in the FQH regime with
imposed geometrical restrictions. This type of devices can be fabricated and
studied experimentally \cite{Leo_1D}. In the first system, an array of gates 
(narrow bridges) is imposed on the strip. We refer to this as case I. The 
second system (case II) is composed of two separate parallel strips connected 
by an array of tunnel barriers. Both systems can be understood as quasi 1D 
arrays of FQH islands connected by tunnel barriers. We include in our model 
the charging energy of the islands in the two cases. 

In the absence of gates, a strip in the FQH regime is circulated by edge 
currents. A gate induces tunneling of quasiparticles 
between the two edges. The result is charge localization-like 
(of an integer charge) 
and Coulomb blockade-like effects. Hence case I is analogous to a 1D array
of serially coupled normal \cite{normal} or Josephson \cite{arrays} junctions. 
In this case we predict the 
existence of integer charge solitons. We discuss two limits according to the
system's parameters: 1. the dissipative limit, in which the solitons behave as
overdamped classical particles; 2. the inertial limit, in which the kinetic
energy becomes important, and the solitons behave as underdamped semiclassical
or quantum particles. 

Case II, in which the charging energy of the islands is included, is analogous 
to a 1D array of Josephson junctions coupled in parallel. A continuous version
of this system, i.e., two parallel strips coupled by a continuous thin tunnel 
barrier (instead of an array of discrete barriers) was considered by 
Wen \cite{Wen_Soliton}. He showed that the system is in the deep 
quantum limit. 
We find that this additional charging energy drives the system towards the 
semi-classical regime, and can produce free solitons.
Again we consider both the dissipative and the inertial limits.

\section{The model}

First we describe the dynamics of edge excitations in the 
FQH samples. As was shown by Wen \cite{Wen_Rev}, 
this dynamics is governed by the following chiral Luttinger liquid Lagrangian:
\begin{equation}
\label{edge_lagrangian}
L = -{\hbar \over 4\pi g} \int dx (\phi_t + v \phi_x) \phi_x \ , 
\end{equation}
where $v$ is the velocity of the edge excitation, $g=\nu$ is the filling 
factor and 
$\phi(x)$ is a 1-D bosonic field, whose physical meaning
is given by the two following relations: $\rho = e\phi_x/{2\pi}$
and $j = -e\phi_t/{2\pi}$. Here $\rho$ is the charge density, while
$j$ is the edge current. Writing the Lagrangian this way, one assumes
that the edge current direction coincides with the direction of the 
x-axis. The model (\ref{edge_lagrangian}) is quantized by assuming the
following commutation relations:
\begin{equation}
\label{comutation_relation_1}
[\phi(x) , \phi(y) ] = {\pi g}\, {\mbox{sgn}}(x-y) \ , 
\end{equation}

\begin{equation}
\label{comutation_relation_2}
[ \pi_{\phi}(x) , \phi(y) ] = {\hbar\over 2} \delta(x-y) \ ,
\end{equation}  
where $\pi_{\phi} \equiv \delta L / \delta \phi_t$.
The remarkable property of the edge excitations that we will be
using later is the fact that the non-zero charge density is always
accompanied by the non-zero current, and vice versa. 

Consider two parallel edges with opposite current
directions (See  Fig.(\ref{FIG_1}a)).
These may be two sides of the same FQH sample or the
two edges of different samples. We will denote these edges by sub-letters
R (right moving) and L (left moving). The R edge is described by the 
Lagrangian (\ref{edge_lagrangian}):
\begin{equation}
\label{right_edge_lagrangian}
L_R = -{\hbar\over 4\pi g} \int dx ({\phi_R}_t+v{\phi_R}_x) {\phi_R}_x \ ,
\end{equation}
while for the L edge the current
direction is opposite to the x-axis direction. Thus, for the L edge:
\begin{equation}
\label{left_edge_lagrangian}
L_L = {\hbar\over 4\pi g} \int dx ({\phi_L}_t-v {\phi_L}_x) {\phi_L}_x \ .
\end{equation}     
The charge density and the current at the L edge are given by
$\rho_L = -e{\phi_L}_x/{2\pi}$ and $j_L = -e{\phi_L}_t/{2\pi}$.
Consider now the two edges as a unified dynamical system.
The net charge density and current of the two edges are then:
\begin{equation}
\label{total_charge}
\rho \equiv \rho_R+\rho_L = {e({\phi_R}_x - {\phi_L}_x) \over 2\pi} \ ,
\end{equation}
\begin{equation}
\label{total_current}
j \equiv j_R+j_L = -{e({\phi_R}_t + {\phi_L}_t) \over 2\pi} \ .
\end{equation} 
The Lagrangian of the combined system is $L=L_R + L_L$. Transforming 
$L$ to the new variables $\phi = \phi_R - \phi_L$, $\theta = \phi_R +
\phi_L$, one arrives at:
\begin{equation}
\label{final_lagrangian}
L = -{\hbar\over 8\pi g}
\int dx (\phi_x\theta_t + \theta_x\phi_t + v\phi_x^2 + v\theta_x^2) \ .
\end{equation}
Introducing two conjugate momenta $\pi_\phi = 2 \delta L / \delta
\phi_t$ and  $\pi_\theta = 2 \delta L / \delta
\theta_t$ (factor 2 is to compensate for the $1/2$ in
(\ref{comutation_relation_2})), one obtains the Hamiltonian in the
$\phi$ representation:
\begin{equation}
\label{hamiltonian_free}
{\cal H}_0 = {\hbar v\over2} 
\int dx \left({\phi_x^2\over 4\pi g} + 4\pi g \pi_\phi^2\right) \ .
\end{equation}
The Hamiltonian (\ref{hamiltonian_free}) corresponds to the usual 
(non chiral)
Luttinger liquid model \cite{Haldane_Lutt}.
The characteristic
energy scale of (\ref{hamiltonian_free}) is the width of the 
band of excitations in the edge channels, 
$\hbar\omega_{\mbox{\footnotesize cut}}$, which is limited by the gap in the 
bulk FQH state. As an upper bound to $\omega_{\mbox{\footnotesize cut}}$ we 
may take the cyclotron frequency. 

We consider now the two systems sketched in Fig.(\ref{FIG_1}b,c).
The first consists of a series of gates imposed on a Hall bar, while
the second is a series of tunnel barriers connecting two FQH bars.
In both systems, between each
pair of neighboring barriers there is an island of mesoscopic size
(taken to be much larger than the width of the barrier). 
We assume, for simplicity, that all the barriers and islands are
equal and we denote by $x_i$ the locations of the barriers.
 
For every barrier in the system we should add to the Hamiltonian
a tunneling term, describing the tunneling of the charge carriers
across the barrier (the back-scattering of the edge
excitations). This term is usually taken as: 
${\cal H}_B \propto \psi^{\dag}_L(x_i) \psi_R(x_i) + h.c.$, 
where $\psi^{\dag}$ is
the creation operator for the tunneling charge carrier.
This term might contain a phase shift brought about by the
coupling of the charge carriers to the external magnetic field
\cite{Fisher_private}. 
In case I the tunneling charge carriers are Laughlin quasiparticles 
(or vortices), thus ${\psi_R}^{\dag} \propto e^{i\phi_R}$ 
(see \cite{Wen_Rev}), and the tunneling Hamiltonian is:  
\begin{equation}
\label{pot_ham_1}
{\cal H}_{B,\,I} = V_{B,}{}_{\,{I}} \sum_i \cos \left[
\phi_R ( x_i ) - \phi_L ( x_i ) \right] = 
V_B \sum_i \cos[\phi(x_i)] \ .
\end{equation}
Here $\phi_i \equiv \phi(x_i) = \phi_R ( x_i ) - \phi_L ( x_i )$
describes the charge that was brought to the barrier at $x_i$.
The tunneling energy scale is the height of the barrier, 
$V_{B,}{}_{\,{I}}\sim\hbar\omega_{\mbox{\footnotesize cut}}|r|^2$,
where $r$ is the reflection amplitude of the barrier.
In this case there is no phase shift in the tunneling Hamiltonian 
(\ref{pot_ham_1}). This may be understood using the 
Ginsburg-Landau-Chern-Simons (GLCS) mean field theory of the FQH effect
\cite{GLCS} or the dual form of this theory \cite{Wen_Niu_90}. The 
vortices in the GLCS theory are topological excitations of a charged 
boson field, which is coupled to a gauge field composed of the external
magnetic field $A$ and a statistical field $a$. The mean field
solution implies that $<A+a> = 0$. The charged vortices thus do not
feel any average gauge field, and do not acquire any phase while moving 
through the FQH liquid. 

In case II the tunneling particles are electrons, 
${\psi_R}^{\dag} \propto e^{i\phi_R/g}$, and the tunneling 
is through potential barriers, where no strongly correlated electron
liquid is present. Therefore the tunneling electrons feel only the 
external magnetic field and the corresponding phase factor should be
taken into account. The tunneling Hamiltonian is thus:
\begin{equation}
\label{pot_ham_2}
{\cal H}_{B,\,II} = 
V_{B,}{}_{\,{II}} \sum_i \cos\left[{\phi(x_i)\over g} + 
{2\pi\Phi_i\over \Phi_0}\right] \ .
\end{equation}
Here, $V_{B,}{}_{\,{II}}$ stands for the interbar tunneling energy 
scale. $\Phi_i$ is the magnetic flux through all the islands (the
places not filled with the FQH liquid) situated to the left of
the tunnel {\it i}th barrier, and $\Phi_0\equiv h/e$ is the 
flux quantum. The fluxes $\Phi_i$ are construction 
dependent. If the magnetic field in the islands is of the same order
of magnitude as the magnetic field in the FQH strips (the most
probable experimental situation) the flux through an island
is very large, so the phase factor in the tunneling Hamiltonian
is random (modulus $2\pi$). One may think, however, of a different 
experimental setup where the fluxes $\Phi_i$ are
under control. Consider, for example, two FQH strips parallel to the 
$x-y$ plane shifted vertically (in the $z$ direction) from each other. 
If the modulations at the edges of the two strips overlap, the tunneling 
between the edges is in the $z$ direction. In such a system all $\Phi_i$'s 
are zero. In what follows we concentrate on this (zero phase shifts) 
situation. The random phase shifts situation resembles the effect of offset
charges in Josephson junction arrays \cite{Fisher_private}, and seems to 
produce a completely different physics. We will consider this situation 
elsewhere.

Next, we should take into account the real charging energy of the
system, caused by local charge fluctuations in the array. 
For a discrete system (like ours), we find it natural to define
the charging energy per island. Thus we add to the Hamiltonian:
\begin{equation}
\label{chrg_ham}
{\cal H}_{C_0}=
\sum_i {E_{C_0}\over (2\pi)^2} [ \phi( x_i ) - \phi ( x_{i+1} ) ]^2 \ , 
\end{equation}
which is the charging energy of the islands. Its energy
scale is $E_{C_0}=e^2/2C_0$, where $C_0$ is either 
the capacitance of an island 
to a substrate or the self capacitance of the island. 
In the first case, this capacitance scales like 
$l$ ($l$ is the length of the island) for a given width of the system. 
In the second case, it scales with the linear size of the island.

So far, we considered three energy scales.
An additional energy scale which plays a role is the energy level
spacing of the edge states within each island, 
$\Delta \epsilon \sim \hbar v / l$.  
We show below that the role of this scale is to determine the crossover
between inertial and dissipative behavior of the system. 

\section{Description in terms of barriers' degrees of freedom}

Following the procedure used in \cite{KF}, we can integrate out 
the degrees of freedom at positions other than the barriers. We thus
obtain an effective Euclidean action in terms of the variable 
$\phi_i$.
The charging energy and the back-scattering terms do not change:
\begin{equation}
\label{coupling_act}
S_{C_0}=\sum_i\int d\tau\,{E_{C_0}\over (2\pi)^2}(\phi_i-\phi_{i+1})^2\ ,
\end{equation}
\begin{equation}
\label{pot_act_1}
{S_{B,\,I}} = -\sum_i\int d\tau\,V_{B,\,{I}}\cos(\phi_i)\ .
\end{equation}
\begin{equation}
\label{pot_act_2}
{S_{B,\,II}}= 
-\sum_i\int d\tau\,V_{B,\,{II}}\cos\left(\phi_i/ g \right)\ .
\end{equation}
In the new coordinates, the self charging and the back-scattering 
energies may be understood as a longitudinal coupling energy 
between the barriers, and as a potential energy, respectively.
The potential energy is in the form of a periodic charging energy 
resulting from non-linear capacitor with capacitance 
$C_{B,}{}_{\,{I}}=e^2/(2\pi)^2 V_{B,\,{I}}$ in case I and
$C_{B,}{}_{\,{II}}=(ge)^2/(2\pi)^2 V_{B,\,{II}}$ in case II.

The integration in the free parts of the Hamiltonian gives:
\begin{eqnarray}
\label{coupling+kin+diss_act}
S_0 & = & {k_BT\over\hbar}\sum_i\sum_n{\hbar v\over 2(4\pi g)}  
\left[{\alpha_n\over \tanh(\alpha_nl)} 
\left(|\phi_{i_n}|^2+|\phi_{i+1_n}|^2\right)\right. \nonumber \\ 
& & \left. -{\alpha_n\over\sinh(\alpha_nl)}
2{\mbox{Re}}(\phi_{i_n}^*\phi_{i+1_n}) \right]\ ,
\end{eqnarray}
where $k_B$ is Boltzman's constant, $\alpha_n\equiv\omega_n/v$, 
$\omega_n$ are Matsubara frequencies, and $\phi_{i_n}$ is a Fourier
component of $\phi(\tau)$. The coefficients of expression 
(\ref{coupling+kin+diss_act}) have a crossover from parabolic to linear 
behavior at a critical frequency $\Omega_{\mbox{\footnotesize cr}}=v/l$.
This critical frequency corresponds to $\Delta\epsilon$. When 
$\omega_1>\Omega_{\mbox{\footnotesize cr}}$, i.e., for long islands and high
temperature, the action (\ref{coupling+kin+diss_act}) becomes dissipative:
\begin{equation}
\label{S_O_diss}
S_{0_{\mbox{\footnotesize diss}}}=
{k_BT\over\hbar}\sum_i\sum_n {\hbar\over 4\pi g}
|\omega_n||\phi_{i_n}|^2\ .
\end{equation}
This dissipation is of the standard, ohmic form\cite{CL}, and it is a 
generalization of the result obtained in \cite{KF}. If, on the other hand,
several $\omega_n$ are in the parabolic section (short islands and low 
temperature), we get instead of dissipation a coupling term 
\begin{equation}
\label{S_0_coupling}
S_{0_C}=\sum_i\int d\tau\,{1\over 2}{\hbar v\over (4\pi g)l}
(\phi_i-\phi_{i+1})^2\ ,
\end{equation}
and a kinetic term
\begin{equation}
\label{S_0_kin}
S_{0_{\mbox{\footnotesize kin}}}=
\sum_i\int d\tau\,{1\over 2}{\hbar l\over (4\pi g)v}
{1\over 3}(\phi_{i,t}^2+\phi_{i+1,t}^2+\phi_{i,t}\phi_{i+1,t})\ .
\end{equation}
If the length of the islands is of the order of $10^3\,$\AA, the temperature
which separating dissipative from inertial behavior is of the order
of $1\,$K.
Regarding the two terms in the inertial limit as a charging and an inductive 
energy terms,
respectively, we can define an internal Hall (or Luttinger) capacitance,
$C_{{H}}=ge^2l/\pi\hbar v$, and an internal Hall (or Luttinger) inductance,
$L_{{H}}=\pi\hbar l/gve^2$. We can express these two properties in a very
simple way by using the propagation time in the grains, $\tau_P=l/v$, and 
the Hall (or Luttinger) resistance, $R_{{H}}= h/ge^2$: 
$C_{{H}}=2\tau_P/R_{{H}}$, and 
$L_{{H}}= \tau_PR_{{H}} /2 $. 
The total charging energy of the system, $\tilde E_{C_0} =e^2/2 \tilde C_0$,
is the sum of the internal and the islands charging energies. The total
capacitance is $\tilde C_0=C_{{H}}/G^2$, where 
$G^2\equiv{1+C_{{H}}/C_0}$.

\begin{figure}
\centerline{\psfig{figure=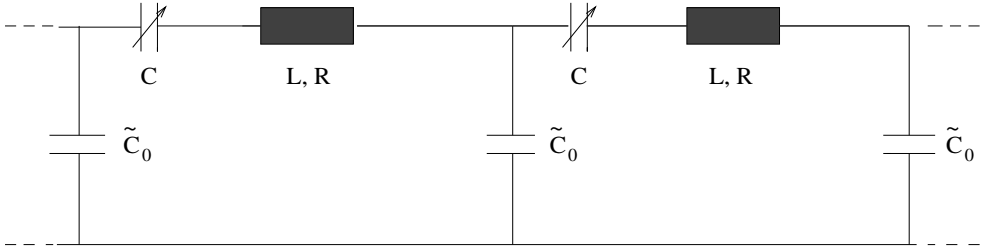,width=13.0cm}}
\caption{Equivalent circuit of the system. The black box
is an inductor in the inertial regime and a resistor in the
dissipative regime.}
\label{FIG_2}
\end{figure}

The above discussion suggests that, under the condition we 
assume here, the array can be represented by the equivalent 
circuit shown in Fig. (\ref{FIG_2}). The non-linear capacitors 
represent the backwards scattering at the barriers, and the linear 
capacitors represent the charging energy of the islands which couples 
neighboring 
barriers. The islands are represented by inductors or by resistors,
according to whether the system is inertial or dissipative, respectively.

In the limit $\tilde E_{C_0} \gg V_B$, the characteristic length scale 
over which the variable $\phi_i$ changes is much larger than the length 
of the unit cell of the array (which is, approximately, the size of an 
island, $l$). If we assume that the fluxes through the islands, $\Phi_i$,
are all zero (constant) in both cases, we can take the continuum limit by 
replacing $( \phi_i - \phi_{i + 1} )^2$ by $ l^2 ( \partial_x \phi )^2$, 
and obtain a pure or an overdamped sine-Gordon model for the inertial or 
the dissipative regimes, respectively. In the inertial limit, the pure 
sine-Gordon Lagrangian is 
\begin{equation}
\label{sine_gordon_1_2}
L_{I,II} = 
{\hbar {v_C}\over 2} {1\over \beta^2_{I,II}}
\int dx  \left\{{1\over v_C^2}\phi^2_t{}_{I,II}- 
\phi^2_x{}_{I,II}- \right. 
\left.  {1\over \Lambda^2_{C,\,{I,II}}}
[1-\cos(\phi_{I,II})]\right\} \ , 
\end{equation} 
where
\begin{equation}
\label{phi_tilde}
\phi_I=g\phi_{II} = \phi  \ ,
\end{equation}

\begin{equation}
\label{beta}
\beta_I^2 = g^2\beta_{II}^2=
{4\pi g \over G^2}\ ,
\end{equation}

\begin{equation}
\label{v_C}
{v_C^2} = v^2 G^2 = {l\over{L_{{H}}\tilde C_0}} \ ,
\end{equation}

\begin{equation}
\label{lambda_I}
\Lambda^2_{C,\,I} = {\hbar v l G^2\over 4\pi g V_{B,\,{I}} }
=l^2{C_{B,}{}_{\,{I}}\over\tilde C_0} \ .
\end{equation}

\begin{equation}
\label{lambda_II}
\Lambda^2_{C,\,II} = {\hbar v l g G^2\over 4\pi V_{B,\,{II}} }
=l^2{C_{B,}{}_{\,{II}}\over\tilde C_0} \ .
\end{equation}
The meaning of the parameters $\beta^2$, $v_C$ and $\Lambda_C$ is explained 
below. We see that in both cases the charging energy of the islands 
renormalizes the parameters of the system.

\section{Charge Solitons}

The two sine-Gordon models presented above admit the existence of topological 
solitons, connecting two adjacent minima of the potential. As $\phi$ is a 
charge field, these are charge solitons. In case I, where the tunneling 
charge is $ge$, the soliton's charge is $e$, as follows from Eq. 
(\ref{total_charge}). The soliton of case I is thus an integer charge soliton. 
It represents the effect of an excess electron in the system, similar to 
charge solitons in arrays of normal \cite{normal} and Josephson \cite{arrays} 
junctions. The number of these solitons can be controlled by introducing a 
gate voltage to one of the islands. However, in contrast to charge solitons in 
arrays of tunnel junctions, the integer charge soliton we study here involves 
current loops in the system, even in a static configuration. As was mentioned 
above, a non-zero charge density must be accompanied by a proportional current
along the edge. Thus to have the inhomogeneous charge density needed for the
soliton, the corresponding edge currents must be partially shortened by tunnel
currents at the barriers. As a consequence, the integer charge soliton carries
a magnetic flux as well. One should not confuse this flux with
the additional external magnetic flux ''occupied'' by the FQH liquid due to
the existence of the soliton. When an electron is injected into the system, 
the incompressible FQH liquid expands in order to maintain its charge density, 
and ''occupies'' an additional external magnetic flux of $\Phi_0/g$. 

In case II, where the tunneling charge is $e$, the soliton's charge is $ge$ 
((\ref{total_charge}), (\ref{phi_tilde})). Therefore it is a {\it fractional}
charge soliton. As such, it can not represent the effect of an external
(integer) charge. In fact it exists in a neutral system. This fractional
charge soliton is basically a Laughlin vortex located between the FQH strips, 
and it can be created by changing the external magnetic field. Thus the 
fractional charge soliton carries a magnetic flux as well, which is equal to 
$\Phi_0$. This soliton can be viewed as an analogue to a fluxon in a 1D array 
of parallely coupled Josephson junctions.

The width of the two kind of solitons, $\Lambda_{C,\,I,II}$ 
((\ref{lambda_I}),(\ref{lambda_II})), is the 
characteristic length scale of the charge density (or current) modulations.
The condition for assuming the continuum limit can thus be written as
$\Lambda_{C,\,I,II}\gg l$ \cite{arrays}. If this condition is not
satisfied, the discreteness nature of the systems should be observed
\cite{Leo_1D}, \cite{discrete}. Here we study only the continuum limit.

In the inertial limit the soliton is a relativistic object with a limiting 
velocity $v_C$ (\ref{v_C}) and a mass
\begin{equation}
\label{mass}
M_{0,\,{I,II}}
={8\over (2\pi)^2}e^2{L_{{H}}\over l}{1\over\Lambda_{C,\,I,II}}\ .
\end{equation}
Both $v_C$ and $M_{0,\,{I,II}}$ depend on $E_{C_0}$. Typically, $v_C$ is 
about two orders of magnitude less than the vacuum light velocity, and 
$M_{0,\,{I,II}}$ is several orders of magnitude less than the electron 
rest mass.

The existence of solitons in the inertial limit depends on the value of the 
coupling constant
$\beta^2$. When $\beta^2= 8\pi$ the ground state of the sine-Gordon model 
becomes unstable \cite{Coleman}. This is a point of a phase transition in
the corresponding $X$-$Y$ model \cite{Samuel}. In the $\beta^2< 8\pi$  
phase, isolated solitons can exist. In the $\beta^2> 8\pi$
phase, solitons and anti-solitons are bound in dipoles.
{}From the expression for $\beta^2$ (\ref{beta}), we notice that for a 
negligible self charging energy ($E_{C_0}=0$) the first system is in the
free solitons phase, while the second one is in the bound solitons phase
(if $g<1/2$). However, when $E_{C_0}$ increases, the value of $\beta^2$ 
decreases. Thus we find that a finite self charging energy drives the system
towards the free solitons phase. System I remains in its initial phase when 
$E_{C_0}$ increases, but system II undergoes a phase transition: when 
$C_0<C_H2g/(1-2g)$, it is in the free solitons phase. 

An exact treatment of the non linear effects, in the presence of
dissipation, is not possible. We can, however, integrate out the high
frequency degrees of freedom, by the methods used, separately, for the
inertial sine-Gordon chain and for the dissipative single Luttinger
junction. The effects of dissipation are formally the same as those
induced by the presence of an infinite number of junctions.
Let us integrate out the modes with frequencies 
$\omega_{cut} - d \omega_{cut}< \omega < \omega_{cut}$, 
lying in the corresponding shell of the wave
numbers $ k $. {}From (\ref{chrg_ham}) and  (\ref{S_O_diss}) 
we find that we have to replace 
$\Lambda^2_{C,\,I,II}$ by:
\begin{equation}
\Lambda '^{-2}_{C,\,I,II} = 
\Lambda^{-2}_{C,\,I,II} e^{- K/2} \ ,
\end{equation}
where:
\begin{equation}
K =\int_{d \omega_{\mbox{\footnotesize cut}}} {d k\over 2\pi} 
{d \omega\over 2\pi}\,
\left[{2E_{C_0}l  k^2\over (2\pi)^2 \hbar } + 
{| \omega |\over 2\pi gl}\right]^{-1} \ .
\label{correl}
\end{equation}
The integral over $\omega$ should be understood as an approximation 
to a sum over Matsubara frequencies in the case of low temperatures.
The double integral in (\ref{correl}) (over $k$ and $\omega$), combined with 
the extra
term in the denominator, implies that the relative strength of the cosine
term (that is, in units of $\omega_{cut}$) always grows, suppressing
the fluctuations in the $\phi_i$'s. Thus, in the presence of
dissipation, solitons are always in the semiclassical regime.

The small mass of the soliton and the absence of dissipation in the 
underdamped 
regime suggests that the soliton can
propagate coherently over the entire array. The soliton can be quantized
semi-classically using collective coordinates, as was done in \cite{arrays}.
Quantum effects of solitons in this regime will be studied elsewhere.

In the overdamped sine-Gordon case (the dissipative regime) one needs to
apply an external voltage in order to obtain propagation of solitons,
i.e., a current. If we assume that this voltage is low enough so that the
soliton is not deformed, the steady state current is proportional to the
voltage with an effective resistance
\begin{equation}
\label{reff}
R_{\mbox{\footnotesize eff},\, {I,II}}=
{8\over(2\pi)^2}{L_{\mbox{\footnotesize tot}}^2
\over\Lambda_{C,\,I,II}l}R_{{H}}\ ,
\end{equation}
where $L_{\mbox{\footnotesize tot}}$ is the length of the array.
If we assume $L_{\mbox{\footnotesize tot}}\approx 10\Lambda_C$,
and $l\approx 10^3$\AA, the effective resistance is larger than the internal
resistance by three orders of magnitude. Here, as well, a future study is
needed.

\section{Conclusions}
We have studied the charge dynamics in arrays of junctions in systems
which are in the fractional
quantum Hall regime. The arrays exhibit quantized charge transport, like
in arrays of low capacitance Josephson or normal tunnel junctions.
The unit of charge, however, can be fractional, reflecting the nature of
the excitations of the system.

The existence of quantized charge solitons makes possible the use of
these devices for the same purposes as the single electron circuits
(electrometers, transistors, turnstiles) widely discussed in the
literature. It is interesting to note that, in the present case,
cotunneling effects are sharply reduced, with respect to devices based
on independent electron tunneling.
The non Fermi nature of the charge carriers  leads to hopping amplitudes
which scale to zero at small frequencies or temperatures. Hence,
coherent transport across two junctions (cotunneling) is suppressed
by extra powers, with respect to the case of normal electrons.

\section{Acknowledgments}

We are thankful to helpful discussions with Y.~Aharonov, N.~Andrei, 
Y. Blanter, L.~Brey, R.~Fazio, G.~G\'omez-Santos, G.~Sch\"on, V.~Yakovenko 
and G.~Zimanyi. We specially thank M.~P.~A.~Fisher for pointing
to us the possibility of phase shifts between tunnel barriers.
We acknowledge financial support from grant 
MAT94-0982 (Spain). E.~B. and A.~S. are supported by Grant from the office 
of the Vice President for research at Tel Aviv University. Z.~H. is supported 
by a MINERVA fellowship.

\baselineskip=1pc

\end{document}